\documentclass{article}
\usepackage{spconf,amsmath,graphicx}
\usepackage{lipsum}  
\usepackage{euscript}
\usepackage{hyperref}

\title{Computational Image-based Stroke Assessment for Evaluation of Cerebroprotectants with Longitudinal and Multi-site Preclinical MRI}
\name{Ryan P. Cabeen$^1$, Joseph Mandeville$^3$, Fahmeed Hyder$^4$, Basavaraju G. Sanganahalli$^4$,  Daniel R. }
\secondlinename{Thedens$^5$, Ali S. Arbab$^6$, Shuning Huang$^7$, Adnan Bibic$^8$, Erendiz Tarakci$^1$, Jelena Mihailovic$^4$,} 
\thirdlinename{Andreia Morais$^3$, Jessica Lamb$^2$, Karisma Nagarkatti$^2$, Arthur W. Toga$^1$, Patrick Lyden$^2$, Cenk Ayata$^3$}

\address{
\small $^1$USC Stevens Neuroimaging and Informatics Institute, $^2$ Zilkha Neurogenetic Institute, Keck School of Medicine of USC; \\
\small $^3$Martinos Center for Biomedical Imaging, Massachusetts General Hospital, $^4$Departments of Radiology and Biomedical \\
\small Imaging, Yale University, $^5$Carver College of Medicine Epidemiology, University of Iowa, $^6$Medical College of Georgia, \\
\small Augusta University, $^7$McGovern Medical School, University of Texas, $^8$Kennedy Krieger Institute, Johns Hopkins University}

\begin{document}

\maketitle

\begin{abstract}

While ischemic stroke is a leading cause of death worldwide, there has been
little success translating putative cerebroprotectants from rodent preclinical
trials to human patients. We investigated computational image-based assessment
tools for practical improvement of the quality, scalability, and outlook for
large scale preclinical screening for potential therapeutic interventions in
rodent models. We developed, evaluated, and deployed a pipeline for image-based
stroke outcome quantification for the Stroke Preclinical Assessment Network
(SPAN), a multi-site, multi-arm, multi-stage study evaluating a suite of
cerebroprotectant interventions. Our fully automated pipeline combines
state-of-the-art algorithmic and data analytic approaches to assess stroke
outcomes from multi-parameter MRI data collected longitudinally from a rodent
model of middle cerebral artery occlusion (MCAO), including measures of infarct
volume, brain atrophy, midline shift, and data quality. We applied our approach
to 1,368 scans and report population level results of lesion extent and
longitudinal changes from injury. We validated our system by comparison with
both manual annotations of coronal MRI slices and tissue sections from the same
brain, using crowdsourcing from blinded stroke experts from the network. Our
results demonstrate the efficacy and robustness of our image-based stroke
assessments.  The pipeline may provide a promising resource for ongoing rodent
preclinical studies conducted by SPAN and other networks in the future.

\end{abstract}


\section{Introduction}

Many proposed stroke treatments have reached clinical trials with preclinical
support from animal models, but few successfully translate to
cerebroprotectants in patients. Technical and procedural challenges contribute
to the failures of preclinical translation, particularly reliability and
reproducibility \cite{bosetti2017translational} \cite{cabeen2020computational}.
We developed a fully automated image analysis pipeline for the Stroke
Preclinical Assessment Network (SPAN) \cite{lyden2022stroke} to help meet these
needs and address critical issues of rigor, transparency, and reproducibility.
The network includes six research universities and a coordinating center (CC)
who manage enrollment of animals, experimental stroke, and blinded and
randomized treatment with several candidate cerebroprotectants.  SPAN examines
both behavioral and tissue readouts of stroke outcome, and we focus here on the
magnetic resonance imaging (MRI) tissue readouts. 

Most preclinical stroke studies measure tissue outcomes with
triphenyltetrazolium chloride (TTC) stained brain sections \cite{Cole1990-hk},
which has several weaknesses: morphometric changes with tissue handling,
variation in preparation and staining intensity across labs, availability at
only a single time point, limited options for subsequent tissue analysis, and
high inter-rater variability \cite{Friedlander2017-fh}. MRI instead enables:
(i) a direct translational path to human diagnostics with multiple biological
and clinical readouts, (ii) repeated longitudinal scans and standardization
across sites, and (iii) preserved brain morphology. Numerous algorithms have
been proposed for MRI-based stroke imaging \cite{Chang2021-hq, Valverde2019-wm,
Chang2020-jr}, demonstrating feasibility of MRI in preclinical stroke studies.

% \cite{Ozertem2007-ux} \cite{Chang2021-hq} \cite{Gupta2014-qq}
% \cite{Mulder2017-lk} \cite{koch2019atlas}, \cite{Valverde2019-wm}
% \cite{Chang2020-jr}

We build on this work to develop a fully automated pipeline for image-based
stroke assessment for SPAN to provide robust processing and
continuous reporting of data from multiple time points after injury at multiple
imaging centers.  We focus on three distinguishing contributions: first,
providing a robust end-to-end solution including measures of lesion volume,
brain atrophy, ventricular volume, midline shift, and data quality; second,
robust handling of multi-site longitudinal data; third, a dataset larger than
any previous preclinical stroke MRI study (N = 1,368). We validated
our approach with blinded and randomized human labeling of TTC tissue staining
and human labeling of MRI sections.

\section{Methods}

\begin{figure*}[!t]
  \centering
  \includegraphics[width=0.9 \textwidth]{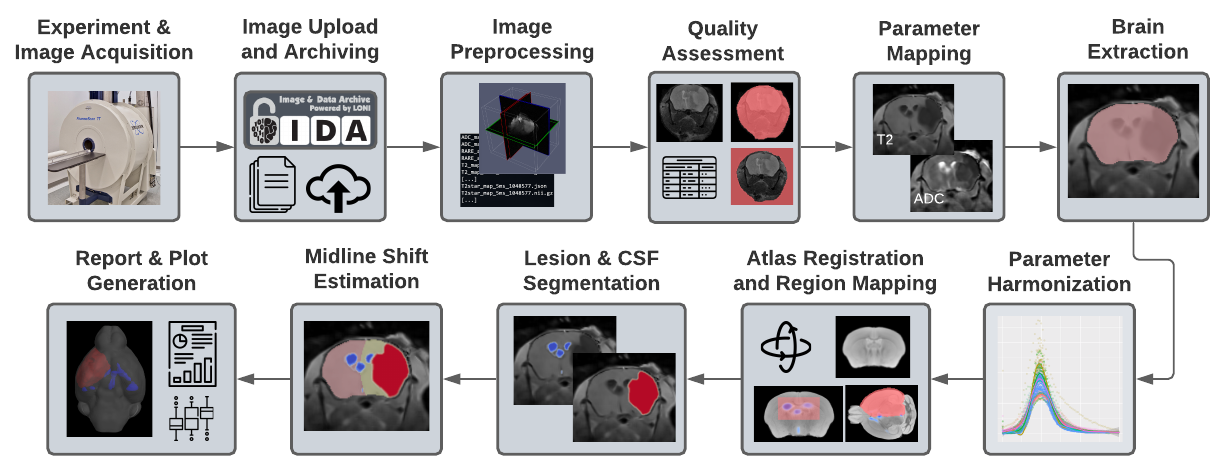}
  \caption{A schematic overview of our image analysis pipeline.}
  \label{overview}
\end{figure*}

Our image analysis software pipeline (Fig. \ref{overview}) was implemented with
the Quantitative Imaging Toolkit (QIT) \cite{cabeen2018quantitative}, Advanced
Normalization Tools (ANTs) \cite{avants2009advanced}, and R 4.1.0 for plotting
and statistical analysis.  The pipeline includes image acquisition,
preprocessing, quality assessment, harmonization, brain and lesion
segmentation, midline shift quantification, and analytic reporting, described
as follows.

{\bf Imaging protocol and data collection:} Data were collected from a mouse
model with experimental middle cerebral artery occlusion (MCAO) at day 2 and
day 30 after injury. With ethics board approval, imaging was performed across
six imaging centers on Bruker scanners (field strengths including 7T, 9.4T,
11.7T); one site used a surface coil and others used a volume coil. The
multi-parameter imaging protocol included multi-echo T2, and diffusion-weighted
MRI (DWI) collected at 150 $\mu$m$^2$ coronal in-plane resolution and 500
$\mu$m slice thickness. All sites used three b-values for DWI (0, 500, 1000
s/mm$^2$) and the T2 protocol used either three echoes (15, 30, 45 ms) or ten
echoes (equally spaced from 10 to 100 ms).  100 mice were scanned in an initial
pilot phase of SPAN, the first stage proceeded to acquire MRI data from 780
animals with a total of 1,368 scanning session, accounting for mortality after
injury.  All data were routinely uploaded by each site in the DICOM format to
the LONI Image Database Archive for long term storage
and analytics.

{\bf Pre-processing and quality assessment:} Data preprocessing included
parsing DICOM tags, sorting by imaging parameters, fixing image coordinates,
converting using dcm2nii, and finally producing a set of matching NIfTI files
for each case. We applied adaptive non-local means denoising with voxelwise
noise estimation, and uniform tricubic resampling at 150 $\mu$m isotropic
resolution. We performed image quality assessment by segmenting foreground and
background using Otsu thresholding and computing the signal-to-noise ratio,
contrast-to-noise ratio, and signal variance-to-noise ratio. We then performed
relaxometry to derive quantitative parameter maps, which included a signal
baseline and rate of decay (T2, ADC) for the multi-echo T2 and DWI scans (Figs.
\ref{results}.A1,\ref{results}.A2) For simplicity of presentation, we report
all T2 values as the inverse relaxation rate (R2).

{\bf Brain segmentation and spatial normalization:} We performed brain
extraction using a deep learning neural network with a U-net architecture (Fig.
\ref{results}.A3) implemented in PyTorch \cite{ronneberger2015u} and similarly
to previous work \cite{Hsu2020-ku} \cite{De_Feo2021-xy}.  We bootstrapped our
model using a semi-automated conventional brain extraction approach applied to
the baseline ADC (least lesion contrast); this process involved edge-preserving
smoothing, gradient computation, thresholding, and morphological operations to
isolate the brain.  We selected training examples from 180 cases and hold-out
testing and validation examples from 30 cases; each was split roughly evenly
between imaging centers and time points. Our network took all four parameter
maps as input (128x128x4 resolution), used a kernel size of 64 and a batch size
of 20, with augmentation by translation, rotation, scaling, contrast, and
deformation. We trained a single 2D U-net with data from each image plane, and
then at inference time, we applied the model to each image slice direction and
computed the average prediction. We trained for 10 epochs on an Nvidia 1080 Ti
12GB GPU for two hours and ten minutes. Following this, we performed linear
registration of each case to the Mouse BIRN Atlasing Toolkit (MBAT) 
\cite{mackenzie2007multimodal} atlas using the T2 rate parameter map.

{\bf Harmonization and lesion segmentation: } There were inter-site differences
in quantiative MRI parameters (Figs. \ref{results}.B1, \ref{results}B3), likely
due to variety in imaging hardware and physiological factors, so we performed
global intensity harmonization of each individual scan (Figs.
\ref{results}.B2, \ref{results}.B4). Using a smoothed histogram of the brain
intensity distribution, we identified the peak value (the mode) and scaled the
entire image to bring the most likely value to one. We chose the mode because
it is less affected by distributional skew due to lesion. We performed lesion
segmentation using multiple thresholding of the harmonized parameter maps
(Fig. \ref{results}.B4). We first defined an initial lesion map by applying an
inverted sigmoidal soft threshold of 0.8 to the R2 map and a threshold
of 1.5 to the ADC map.  We applied a median smoothing filter and a
hysteresis threshold to extract a lesion mask, with a strong threshold of 0.55
and a weak threshold of 0.45. We performed morphological opening to refine the
mask and reduce spurious voxel labels. We then applied an atlas-defined
restriction mask to exclude lesion labels on the contralateral side to injury.
We applied a similar procedure to segment cerebrospinal fluid (CSF) with a
R2 threshold of 0.75 and ADC threshold of 1.25 (not
inverted). Hence, we identified lesion as areas with both dark T2 and ADC, and
CSF as areas with dark R2 and bright ADC.  We then took the remainder of the
brain to be a third segment for ``normal appearing tissue''. This thresholding
approach was jointly chosen by the network’s imaging team with the strict goal
of having an interpretable and understandable lesion definition.

{\bf Midline shift quantification:} We estimated midline shift metrics
including the raw lateral displacement, a normalized shift index, and ratio of
hemispheric volumes.  We first estimated the anatomical midpoint based on
ventricular geometry, using an atlas-defined restriction mask to select voxels
from the lateral and third ventricles in a coronal section (seven voxels thick)
located roughly at the midpoint of the corpus callosum and computed the average
3D CSF position. We defined a surface splitting the hemispheres based on this
estimated midpoint and fit an implicit quadratic surface based on the extreme
points. The surface evaluated to produce volumetric masks for each hemisphere
(Fig.  \ref{results}.B4).  We computed the absolute midline shift from the
difference between the estimated midline and the typical atlas midline. We
computed a normalized midline shift index from the ratio of the absolute shift
over the width of the brain, as well as left and right hemisphere volumes and
their lateralization index.

{\bf Analytic reporting and network feedback: } We finally created reports
summarizing these outcome measures across the study cohort.  This included 3D
visualizations (Figs. \ref{results}.A.5, \ref{results}.A.6) and mosaic plots
showing a matrix of coronal sections with segmentation masks superimposed. We
also created data tables which are shared biweekly with the statistics team.
Data was processed at the USC Neuro Imaging Computing Center (NICC) using a
4096 core Sun Grid Engine computing environment.

\section{Results and Discussion}

\begin{figure*}[!t]
  \centering
  \includegraphics[width=0.9\textwidth]{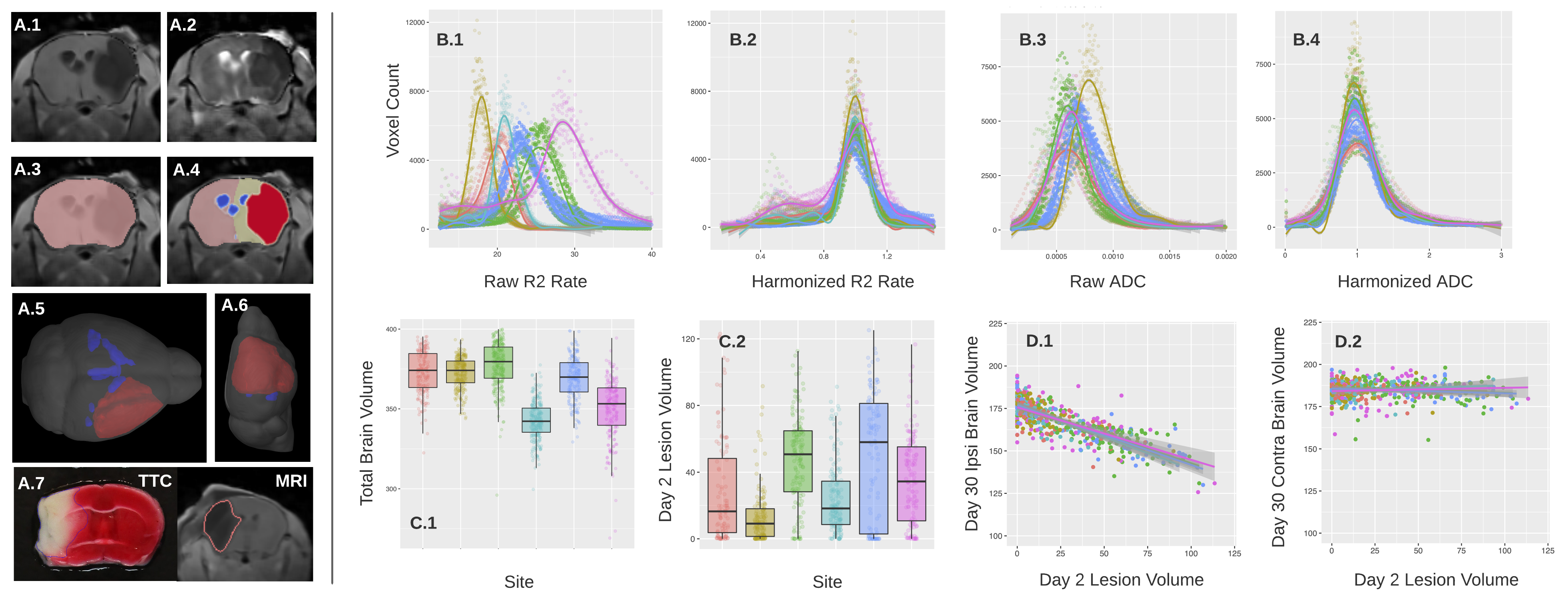}
  \caption{Results of the analysis of SPAN data including multi-parameter MRI (A.1, A.2), segmentation results (A.3-6), validation with TTC (A.7), parameter harmonization (B), group level volumetry (C), and longitudinal effects separated by hemisphere and covaried by site (D). The plots show data from six imaging centers, which are anonymized but colored uniquely.}
  \label{results}
\end{figure*}

{\bf System refinement and testing:}  Our final brain extraction model had a
Dice score of 0.964 on the held-out test dataset. We found that T2 parameters
had greater inter-site variability than ADC; however, both modalities required
harmonization for fully automated lesion segmentation (Fig. \ref{results}.B.2,
\ref{results}.B.4).  We found that several sites had image artifact due to the
MCAO catheter and also the imaging hardware; these were sometimes misclassified
as lesion, but this could be simply excluded by the atlas lesion restriction
mask.  We tested a range of lesion and CSF thresholds within $\pm$ 0.05 of our
chosen thresholds and found results to be robust to minor changes. We analyzed
a total of 1,368 scans, and among these there were 20 cases excluded, including
one due to file transfer issues, six due to missing scans, and 13 due to
excessive motion.  A final count of 1,348 scans passed QC with an analytics
completion rate of 98.5\%. The average time to process one case was one hour
and 48 minutes.

{\bf Group-level analysis of lesion extent and longitudinal changes:}  We
computed distributional statistics of total brain volume, lesion volume,
midline shift, and atrophy in aggregate and split by site (Figs.
\ref{results}.C, \ref{results}.D).  We also computed lesion probability maps
for the entire cohort and split by site; these were visualized on sectional
anatomy of the atlas and also as 3D surface renderings (Figs.
\ref{results}.A.5, \ref{results}.A.6).  We examined longitudinal changes in
outcome measures by comparing day 2 lesion volume to day 30 atrophy (both whole
brain and split by hemisphere).  The results indicate that there exists a small
but consistent differences in apparent total brain volume by site (F$_{5,
1347}$ = 264.4, p $<$ 10$^{-15}$), suggesting that per-site corrections should
be included in statistical models.  We found inter-site variability in lesion
volume, but the anatomical location of lesion was consistent across sites.  The
average lesion volume was 32.26 mm$^{3}$.  With site as a covariate, we found
that day 2 volume was highly predictive of hemispheric atrophy on the side
ipsilateral to the injury at day 30 (R$^2$ = 0.73, $\beta$ = -0.34, p $<$
10$^{-15}$; Fig. \ref{results}.D.1), while there was no significant effect
contralateral to injury at day 30 (R$^2$ = 0.57, $\beta$ = -0.01, p = 0.23;
Fig. \ref{results}.D.2).  This supports the general understanding that MCAO
later leads to localized tissue atrophy, and it supports the notion that the
pipeline is able to measure both lesion and lateralized atrophy with high
fidelity.

{\bf Validation with manual tracing of MRI sections:} We performed an initial
validation experiment to determine the accuracy of our lesion segmentation. We
first selected ten typical day 2 scans, and two SPAN network stroke experts
estimated lesion volume from manually delineated coronal
sections of the R2 parameter map using ImageJ.  All ten cases were
analyzed by both raters, and they were blinded to the results of our pipeline.
The same ten cases were subsequently processed with our pipeline without manual
intervention to obtain the lesion volume for comparison in each case.  We
compared the human raters and the automated approach by computing descriptive
statistics, the root-mean-square error (RMSE), and Pearson's correlation
coefficient.  We found the average lesion volume from manual raters was 11.56
mL and from the automated approach was 11.38 mL.  The RMSE error between the
two human raters was 2.22 mL, and the RMSE error between the humans and
automated approach was 2.99 mL.  The correlation between the human raters was
0.969 and that between the humans and automated approach was 0.957.  The
results indicate that the automated approach recovered similar lesion estimates
as human raters, and difference in performance was practically the same as
between humans.

{\bf Validation with manual tracing of TTC stained sections:} We performed a
validation experiment to address whether the MRI-derived lesion metrics reflect
tissue-level changes as observed using the standard approach of TTC stained
sections (Fig. \ref{results}.A7).  For this, 37 mice were randomly selected,
treated with MCAO, and imaged at day 2. But unlike others enrolled in the
study, these cases were sacrificed immediately after imaging and their brains
were subsequently sectioned in 2 mm thick slices, stained with TTC,
photographed on slides (front and back), and the images were shared with the
imaging team.  The cases were split among the sites to ensure the TTC stains
were representative of the variation in the greater stroke field.  This
resulted in a total of 746 individual images, and we chose a total of six
annotators (one from each site) to provide at least two annotation for each
image.  Due the scale and logistics of the task, we built an online web-based
annotation system based on LabelMe \cite{russell2008labelme} to support remote
annotations.  We modified LabelMe to use a restricted set of drawing
operations, to anonymize the images and raters, to enable deployment across
multiple isolated servers for each rater, and to provide instructions to label
the brain, lesion, and any other relevant features.  Once collected, we
computed the lesion volume fraction from the ratio of the total lesion area
over the total brain surface area.  The MRI data from these 37 were processed
like the other cases without manual intervention, and the lesion volume
fraction was computed similarly.  We estimated the reliability among human-TTC
raters using the coefficient of variation (CoV = $\sigma$ / $\mu$) and measured
Pearson's correlation coefficient between the human-TCC and automated-MRI
lesion volume fractions.  We found human-TTC raters to be highly reliable in
delineating the brain (CoV = 2.50\%); however, lesion segmentations had greater
variability (CoV = 18.5\%).  Given this, we also selected a subgroup of
``high-reliability'' TTC cases (N= 24 cases with CoV $<$ 5\%).  The overall
Pearson correlation between the automated-MRI volume fraction and the human-TTC
in all cases was 0.743, and in the ``high-reliability'' cases, the correlation
was 0.865.  The results generally indicate the substantial agreement between
lesion quantification gathered from both TTC tissue and automated MRI analysis.
TTC staining is considered the gold-standard technique for stroke preclinical
imaging, but even so, inter-rater variability in several cases demonstrated
potential limitations as well.  One explanatory factor may be temporal
effects, as the experiment requires some time to elapse between MR imaging
and tissue staining. 

{\bf Conclusions:} We developed, evaluated, and deployed a pipeline for
image-based stroke outcome quantification for SPAN that combines
state-of-the-art data analytic approaches to assess stroke outcomes from
preclinical multi-parameter MRI data collected longitudinally from a rodent
model of middle cerebral artery occlusion (MCAO), including measures of lesion
extent, brain atrophy, midline shift, and data quality.  We applied our
approach with the largest preclinical cohort of mice to date and rigorously
evaluated the validity of our approach with expert manual annotations of MRI
and TTC-stained sections from the same specimens, with the results suggesting
the efficacy and robustness of the pipeline. Open challenges include increasing
the imaging resolution, inclusion of additional modalities for detecting
hemorrhage and water content, and combined analysis with behavioral readouts.
Our pipeline may provide a resource for ongoing preclinical studies conducted
by SPAN and others in the future, and our software may be found online
\cite{Code}.

\newpage

\noindent {\bf Acknowledgements:} Supported by National Institutes of Health
(NIH) grant NS U24 NS113452 from the National Institute of Neurological
Disorders and Stroke (NINDS) (PI: Author PL). RPC Supported by the CZI Imaging
Scientist Award Program number 2020-225670 from the Chan Zuckerberg Initiative
DAF, an advised fund of Silicon Valley Community Foundation (PI: Author RPC).

\bibliographystyle{IEEEbib}
\bibliography{paper}

\end{document}